\documentclass[pra,aps,twocolumn,showpacs]{revtex4}
\usepackage{amssymb}
\usepackage{amsmath}
\usepackage[dvips]{graphicx}
\usepackage[english]{babel}

\begin{document}

\title{Vortex structure in spinor F=2 Bose-Einstein condensates}
\author{W. V. Pogosov, R. Kawate, T. Mizushima, and K. Machida}
\affiliation{Department of Physics, Okayama University, Okayama 700-8530, Japan}
\date{\today }

\begin{abstract}
Extended Gross-Pitaevskii equations for the rotating $F=2$ condensate in a
harmonic trap are solved both numerically and variationally using trial
functions for each component of the wave function. Axially-symmetric vortex
solutions are analyzed and energies of polar and cyclic states are
calculated. The equilibrium transitions between different phases with
changing of the magnetization are studied. We show that at high
magnetization the ground state of the system is determined by interaction in
"density" channel, and at low magnetization spin interactions play a
dominant role. Although there are five hyperfine states, all the particles
are always condensed in one, two or three states. Two novel types of vortex
structures are also discussed.




\end{abstract}

\maketitle


\section{Introduction}

Properties of Bose-Einstein condensates (BEC) of alkali atom gases attract a
considerable current interest. Recently quantized vortices and lattices of
vortices have been obtained experimentally in BEC clouds confined by
magnetic traps \cite{1,2,3,4}. BEC's can have internal degrees of freedom
associated with the hyperfine spin. Such condensates are usually called
spinor BEC's. First examples of these systems with hyperfine spin $F=1$ were
found in optically trapped $^{23}$Na \cite{Stenger}. In zero magnetic field,
spin $F=1$ condensate can be in two different states, which are called
ferromagnetic and polar \cite{Machida,Ho}. Depending on the values of
interaction parameters, which determine coupling between different hyperfine
states, one of these states has a lowest energy. Vortex matter in spinor
BEC's is represented by a rich variety of rather exotic topological
excitations. These vortices were investigated in a large number of
theoretical works for the case $F=1$ (see, e.g., Refs. \cite%
{Mizushima1,Yip,Isoshima1,Isoshima2,Mizushima2,Reijnders,Martikainen,Kita,Mueller,Anglin}%
).

In the most recent experiments, $F=2$ spinor Bose-Einstein condensates have
been created and studied \cite{Schmal,Chang,Gorlitz,Kuwamoto}. However,
superfluid phases in the $F=2$ BEC were analyzed only for the case of the
absence of magnetic field and rotation \cite{Ciobanu} (see also Refs. \cite%
{Ueda1,Ueda2}). Spinor $F=2$ BEC has one more interaction parameter as
compared to the $F=1$ BEC because of the larger spin value. Therefore, there
are three possible phases in absence of magnetic field: polar, ferromagnetic
and cyclic states.

Due to the internal degree of freedom, a rich variety of exotic vortices
have been proposed in $F=1$ spinor BEC's by a large number of authors \cite%
{Kasamatsu2005}. For instance, $F\!=\!1$ spinor BEC's with the ferromagnetic
spin interaction exhibit SO(3) symmetry in spin space, which means that the
local spins may sweep the whole or half the unit sphere. It has been found
that this topological excitation, called the Mermin-Ho vortex, can be
stabilized in the rotating system \cite{Mizushima1}. In the case of $F=2$
spinor BEC's, the possibility of such the coreless vortex state has been
predicted by Zhang \textit{et al.} \cite{Zhang}. However, in the possible
kinds of atoms, such as $^{87}$Rb and $^{23}$Na, the estimated spin
interactions are situated in the close vicinity of the phase boundary
between polar and cyclic phases \cite{Ciobanu}; The detailed study on the
rotating ground state with the cyclic spin interaction is an unexplored
region.

The aim of the present work is to study vortex structure in rotating spinor $%
F=2$ condensate having finite magnetization. The condensate wave function
has five components. We solved the extended Gross-Pitaevskii equations both
numerically and variationally using trial functions for each component of
the wave function. There is a good agreement between the results of both
methods. We restricted our consideration only to the case of
axially-symmetric solutions. Energies of polar, ferromagnetic and cyclic
states with various sets of winding numbers for different components of the
order parameter were evaluated. The equilibrium transitions between
different phases with changing of the magnetization were studied.

\section{Theoretical formalism}

Consider two-dimensional $F=2$ condensate with $N$ particles confined by the
harmonic trapping potential 
\begin{equation}
U(r)=\frac{m\omega _{\perp }^{2}r^{2}}{2},  \label{poten}
\end{equation}%
where $\omega _{\perp }$ is a trapping frequency, $m$ is the mass of the
atom, and $r$ is the radial coordinate. The system is rotated with the
angular velocity $\Omega $. The energy of the system depends on three
interaction parameters $\alpha ,$ $\beta ,$ and $\gamma ,$ which can be
defined as \cite{Ciobanu}%
\begin{eqnarray}
&&\alpha =\frac{1}{7}(4g_{2}+3g_{4}),  \label{alfa_def} \\
&&\beta =-\frac{1}{7}(g_{2}-g_{4}),  \label{bita_def} \\
&&\gamma =\frac{1}{5}(g_{0}-g_{4})-\frac{2}{7}(g_{2}-g_{4}),
\label{gamma_def}
\end{eqnarray}%
where ($q=0,2,4$) 
\begin{equation}
g_{q}=\frac{4\pi \hbar ^{2}}{m}a_{q}  \label{inter_const_def}
\end{equation}%
and $a_{q}$ is the scattering lengths characterizing collisions between
atoms with the total spin $0$, $2$, and $4$.

The order parameter in $F=2$ case has five components $\Psi _{i}$ $%
(i=-2,-1,0,1,2)$. The free energy of the system can be written as \cite%
{Machida,Ho} 
\begin{eqnarray}
F &=&\int dS\left[ \Psi _{j}^{\ast }\widehat{h}\Psi _{j}+\frac{\alpha }{2}%
\Psi _{j}^{\ast }\Psi _{k}^{\ast }\Psi _{j}\Psi _{k}\right.   \notag \\
&&+\frac{\beta }{2}\Psi _{j}^{\ast }\Psi _{l}^{\ast
}(F_{a})_{jk}(F_{a})_{lm}\Psi _{k}\Psi _{m}  \notag \\
&&+\frac{\gamma }{2}\Psi _{j}^{\ast }\Psi _{k}^{\ast }\Psi _{-j}\Psi
_{-k}(-1)^{j}(-1)^{k}  \notag \\
&&\left. -B_{z}M-i\hbar \mathbf{\Omega \cdot }\Psi _{j}^{\ast }(\mathbf{%
\nabla \times r})\Psi _{j}\right] ,  \label{energy_def}
\end{eqnarray}%
%
%
%
%
%
%
%
%
where integration is performed over the system area, repeated indices are
summed, $B_{z}$ is the magnetic field, which is treated as a Lagrange
multiplier, $\widehat{h}$ and $M$ are the one-body Hamiltonian and
magnetization, which are given by 
\begin{eqnarray}
&&\widehat{h}=-\frac{\hbar ^{2}\nabla ^{2}}{2m}+U(r),  \label{one-hamil} \\
&&M=\int dS\left\vert \Psi _{i}\right\vert ^{2}i.  \label{magnetiz_def}
\end{eqnarray}%
Here $F_{a}$ $(a=x,y,z)$ is the angular momentum operator and it can be
expressed in a matrix form as 
\begin{eqnarray}
&&F_{x}=\frac{1}{2}\left( 
\begin{array}{ccccc}
0 & 2 & 0 & 0 & 0 \\ 
2 & 0 & \sqrt{6} & 0 & 0 \\ 
0 & \sqrt{6} & 0 & \sqrt{6} & 0 \\ 
0 & 0 & \sqrt{6} & 0 & 2 \\ 
0 & 0 & 0 & 2 & 0%
\end{array}%
\right) ,  \label{Fx} \\
&&F_{y}=\frac{i}{2}\left( 
\begin{array}{ccccc}
0 & -2 & 0 & 0 & 0 \\ 
2 & 0 & -\sqrt{6} & 0 & 0 \\ 
0 & \sqrt{6} & 0 & -\sqrt{6} & 0 \\ 
0 & 0 & \sqrt{6} & 0 & -2 \\ 
0 & 0 & 0 & 2 & 0%
\end{array}%
\right) ,  \label{Fy} \\
&&F_{z}=\left( 
\begin{array}{ccccc}
2 & 0 & 0 & 0 & 0 \\ 
0 & 1 & 0 & 0 & 0 \\ 
0 & 0 & 0 & 0 & 0 \\ 
0 & 0 & 0 & -1 & 0 \\ 
0 & 0 & 0 & 0 & -2%
\end{array}%
\right) .  \label{Fz}
\end{eqnarray}

It is convenient to introduce two additional order parameters $\left\langle 
\mathbf{f}\right\rangle =\Psi _{j}^{\ast }\mathbf{F}_{jk}\Psi
_{k}/\left\vert \Psi \right\vert ^{2}$ and $\Theta =(-1)^{j}\Psi _{j}^{\ast
}\Psi _{-j}/\left\vert \Psi \right\vert ^{2}$\ characterizing ferromagnetic
ordering and formation of singlet pairs, respectively \cite{Ciobanu}. In the
absence of magnetic field and rotation, BEC can be in three different states 
\cite{Ciobanu} that is easily seen from Eq.~(\ref{energy_def}). These states
are called ferromagnetic, cyclic and polar \cite{Ciobanu}. In ferromagnetic
phase, only one component of the order parameter is nonzero: $\Psi _{-2}=1.$
In cyclic phase, $\Psi _{-1},\Psi _{1}=0$ and $\Psi _{-2}=\frac{1}{2}%
e^{i\theta },$ $\Psi _{0}=\frac{1}{\sqrt{2}},$ $\Psi _{2}=\frac{1}{2}%
e^{-i\theta },$ where $\theta $ is an arbitrary phase (energy of the system
is degenerate with respect to $\theta $). In polar phase, there are three
possibilities: in the first case $\Psi _{-2}=\frac{1}{\sqrt{2}}e^{i\vartheta
},$ $\Psi _{-1},\Psi _{0},\Psi _{1}=0,$ $\Psi _{2}=\frac{1}{\sqrt{2}}%
e^{i\upsilon },$ in the second case $\Psi _{-1}=\frac{1}{\sqrt{2}}%
e^{i\vartheta },$ $\Psi _{-2},\Psi _{0},\Psi _{2}=0,$ $\Psi _{1}=\frac{1}{%
\sqrt{2}}e^{i\upsilon }$, and in the third case $\Psi _{0}=1$\ and all other
components are equal to zero. Here values of $\vartheta $\ and $\upsilon $\
are arbitrary and the energy is degenerate with respects to them. Depending
on values of scattering lengths $a_{q}$ ferromagnetic, cyclic or polar phase
has the lowest energy \cite{Ciobanu}. In ferromagnetic phase, $\Theta 
\mathbf{=}0\mathbf{,}$\ $\left\vert \left\langle \mathbf{f}\right\rangle
\right\vert =2$; in cyclic phase, $\Theta \mathbf{=}0\mathbf{,}$\ $%
\left\langle \mathbf{f}\right\rangle =0$; in polar phase, $\left\vert \Theta
\right\vert \mathbf{=}1\mathbf{,}$\ $\left\langle \mathbf{f}\right\rangle =0$%
.

\bigskip Extended Gross-Pitaevskii equations can be obtained in a standard
way from the condition of minimum of free energy of the system Eq.~(\ref%
{energy_def}): 
\begin{eqnarray}
&& \{\widehat{h}-\mu +\alpha \Psi _{k}^{\ast }\Psi _{k}\}\Psi _{j}+\beta
\{(F_{\alpha })_{lm}(F_{\alpha })_{jk}\Psi _{l}^{\ast }\Psi _{m}\Psi
_{k}\} \nonumber \\
&& +\gamma (-1)^{j}(-1)^{k}\Psi _{k}^{\ast }\Psi _{k}\Psi _{j} \nonumber \\ 
&& -i\hbar \mathbf{\Omega }\cdot \nabla \times \mathbf{r} \Psi
_{j}-B_{z}j\Psi _{j}=0,  \label{GP_eq}
\end{eqnarray}%
where a chemical potential $\mu $\ is interpreted as the Lagrange
multiplier. We use the total number of particles $N=\int dS\Psi _{i}\Psi
_{i}^{\ast }$\ and the magnetization $M$\ as independent variables. \bigskip

\section{Vortex phases and energy}

\subsection{Classification of phases}

Five nonlinear Gross-Pitaevskii equations Eq.~(\ref{GP_eq}) are coupled and
they can be solved numerically. However, some important consequences can be
derived from the preliminary analysis of these equations. In this paper, we
consider only axially-symmetric solutions of the Gross-Pitaevskii equations.
In this case, each component of the order parameter $\Psi _{j}$\ can be
represented as 
\begin{equation}
\Psi _{j}\left( r,\varphi \right) =f_{j}(r)\exp (-L_{j}\varphi ),
\label{Psi_rad}
\end{equation}%
where $\varphi $ is a polar angle and $L_{j}$\ is a winding number. Axial
symmetry of the solution implies that there are some constraints for the
possible sets of $L_{j}.$ It can be shown from Eq.~(\ref{GP_eq}) that $L_{j}$%
\ obeys the following equations 
\begin{eqnarray}
&&L_{2}+L_{1}+L_{-1}+L_{-2}-4L_{0}=0,  \label{L_rule_1} \\
&&L_{2}+L_{-2}-L_{-1}-L_{1}=0,  \label{L_rule_2} \\
&&L_{2}-2L_{-2}-2L_{1}+2L_{-1}=0.  \label{L_rule_3}
\end{eqnarray}%
Eqs.~(\ref{L_rule_1})-(\ref{L_rule_3}) were obtained under the condition
that all the five components of the order parameter are nonzero. If some of
these components are equal to zero identically then other possibilities
appear for the $L_{j}$\ values. We list here all the possible phases
different from ordinary vortex-free state: $(1,1,1,1,1),$ $(-1,\times
,0,\times ,1),$ $(1,\times ,0,\times ,-1),$ $(\times ,0,\times ,1,\times ),$ 
$(\times ,1,\times ,0,\times ),$ $(0,\times ,1,\times ,2),$ $(2,\times
,1,\times ,0),$ $(\times ,2,\times ,1,\times ),$ $(\times ,1,\times
,2,\times ),$ $(-2,-1,0,1,2),$ $(2,1,0,-1,-2),$ $(-1,0,1,2,3),$ $%
(3,2,1,0,-1),$ $(4,3,2,1,0),$ $(0,1,2,3,4).$ Here numbers denote values of $%
L_{j}$, "$\times $" denotes zero value of the corresponding component of the
order parameter. We restricted ourselves only to the cases, when the largest
winding number does not exceed $4$, vortices with higher winding numbers are
assumed to be nonstable.

\subsection{Method}

Now we can find the solutions of the Gross-Pitaevskii equations for each
phase listed above. For the solution of Eq.~(\ref{GP_eq}), we apply a
numerical method, which was used before in Ref.~\cite{Mizushima3}.

\bigskip Besides the numerical solution, we also use a variational \textit{%
ansatz} based on trial functions for each component of the order parameter.
It follows from Eq. (\ref{GP_eq}) that each component of the order parameter
has an asymptotic $f_{j}(r)\sim r^{L_{j}}$ at $r\rightarrow 0$\ and that in
the expansion of $f_{j}(r)$\ in powers of $r$ there are only terms
proportional to $r^{L_{j}+2n}$, where $n\geq 0$ is an integer number.
Superfluid density vanishes at infinite distances from the center of the
potential well and $f_{j}(r)\rightarrow 0$ at $r\rightarrow \infty $.
Therefore, we have chosen the following trial function 
\begin{equation}
f_{j}(r)=C_{j}\left[ \left( \frac{r}{R_{j}}\right) ^{L_{j}}+a_{j}\left( 
\frac{r}{R_{j}}\right) ^{L_{j}+2}\right] \exp \left( -\frac{r^{2}}{2R_{j}^{2}%
}\right) ,  \label{trial_function_def}
\end{equation}%
where $C_{j}$, $a_{j}$\ and $R_{j}$ are variational parameters. Parameters $%
C_{j}$ are not completely independent since they are related by one equation
that is a condition of equality of number of atoms to the given number.

In the limit of noniteracting gas, Gross-Pitaevskii equation becomes linear
with respect to the order parameter. In this case, it is easy to see that $%
a_{j}=0.$\ In the Thomas-Fermi regime this is no longer valid and the system
tries to minimize its energy by changing parameters $a_{j}$\ and $R_{j}$\ as
compared to the limit of an ideal gas. According to our estimates, for the
case of one-component order parameter, trial function (\ref%
{trial_function_def}) is able to give rather accurate results for the energy
and for the rotation frequency of transition from the vortex-free to the
vortex state even in the 'moderate' Thomas-Fermi limit. Therefore, in this
paper we apply the method to the case of five-component order parameter.
Note that variational approaches were applied before for the analysis of
vortex structures in mesoscopic superconductors within the Ginzburg-Landau
theory, see e.g. \cite{Pogosov}, and for vortices in scalar and spinor BEC 
\cite{Fetter,Ueda3}.

Using Eq. (\ref{energy_def}) and the normalization condition for the order
parameter one can find the energy of the system analytically as a function
of variational parameters for each set of $L_{j}$. However, final expression
for the energy is rather cumbersome and we do not present it here. Values of
variational parameters can be calculated by a numerical minimization of the
energy.

\subsection{Results and discussion}

We consider the situation, when the concentration of atoms in $z$-direction
is equal to $2000$ $\mu \mathrm{m}^{-1}$, and the scattering length $a_{0}$
equals $5.5$ $\mathrm{nm}$. According to the estimates made in Ref.~\cite%
{Ciobanu}, $^{23}$Na, $^{83}$Rb, $^{87}$Rb, $^{85}$Rb correspond to points
on the phase diagram in $a_{2}-a_{4}$ vs. $a_{0}-a_{4}$ plane (in absence of
magnetic field and rotation), which are situated in the close vicinity to
the phase boundaries between ferromagnetic, polar and cyclic states (see
Fig.~1 in Ref.~\cite{Ciobanu}). There are even some uncertainties in
positions of these points on the phase diagram because of the error bars in $%
a_{q}$. It was assumed in Ref.~\cite{Ciobanu} that $^{23}$Na, $^{83}$Rb, and 
$^{85}$Rb BEC's are in polar, ferromagnetic, and cyclic states,
respectively, and $^{87}$Rb corresponds to the phase boundary between the
polar and cyclic states. In this paper, we perform all the calculations for
the polar state at $\beta =\frac{\alpha }{50}$, $\gamma =-\frac{\alpha }{50}$%
. For the cyclic state we put $\beta =\frac{\alpha }{50}$, $\gamma =\frac{%
\alpha }{50}$. And for the state situated on the phase boundary between
cyclic and polar state, which we call cyclic+polar, we use $\beta =\frac{%
\alpha }{50}$, $\gamma =0$. We calculated the dependences of the energy of
the system on the magnetization for these three phases for different vortex
states, when the system is rotated with the frequency $\Omega =0.4\omega
_{\perp }$. Our results obtained by the numerical solution to the
Gross-Pitaevskii equations are presented on Fig.~1 for cyclic (a), polar
(b), and cyclic+polar (c) states. In Table 1 we show numerically and
variationally calculated values of the magnetization, at which the
transitions occur between different phases for the case of cyclic state
(Fig. 1 (a)). In this table, "a" denotes the transition between $%
(-1,0,1,2,3) $\ and $(\times ,0,\times ,1,\times )$\ states, "b" between $%
(\times ,0,\times ,1,\times )$\ and $(-2,-1,0,1,2)$\ states, and "c" between 
$(-2,-1,0,1,2)$\ and $(-1,\times ,0,\times ,1)$\ states. There is a good
agreement between the numerical and variational results.

\begin{table}[bp]
\caption{ Values of the magnetization corresponding to the transitions
between different ground states for the case of cyclic phase (Fig. 1(a)),
which were calculated numerically and variationally. }%
\begin{ruledtabular}
\begin{tabular}{cccc}
 & $a$ & $b$ & $c$ \\
\hline
numerical & $0.34$ & $0.79$ & $1.36$ \\ 
variational & $0.38$ & $0.76$ & $1.40$
\end{tabular}
\end{ruledtabular}
\end{table}

In general case, there can be phase differences between functions $f_{j}(r)$%
.\ Axial symmetry of the solution implies that there are some constraints on
phases, which are similar for the constraints on winding numbers and can be
also obtained from the GP equation (\ref{GP_eq}). The energy of the system
is then degenerate with respect to remaining phases, as in the nonrotating
case, which was discussed above.

In all states under study, it also turns out that for the condensate it is
energetically favorable to be distributed between one, two of three
hyperfine states and not between four or five.

It can be seen from Fig.~1 that, in all cases, at zero magnetization, states
with nonzero winding numbers are energetically favorable due to the rotation
of the system. By changing the magnetization it is possible to jump from one
vortex phase to another one. The transitions between different phases are
discontinuous. Note that in this paper we consider only the thermodynamical
transitions between different states. Actual position of transition line
between different vortex phases is controlled by local stability of states
and, therefore, the prehistory of the system.

\begin{figure}[b]
\includegraphics[width=0.8\linewidth]{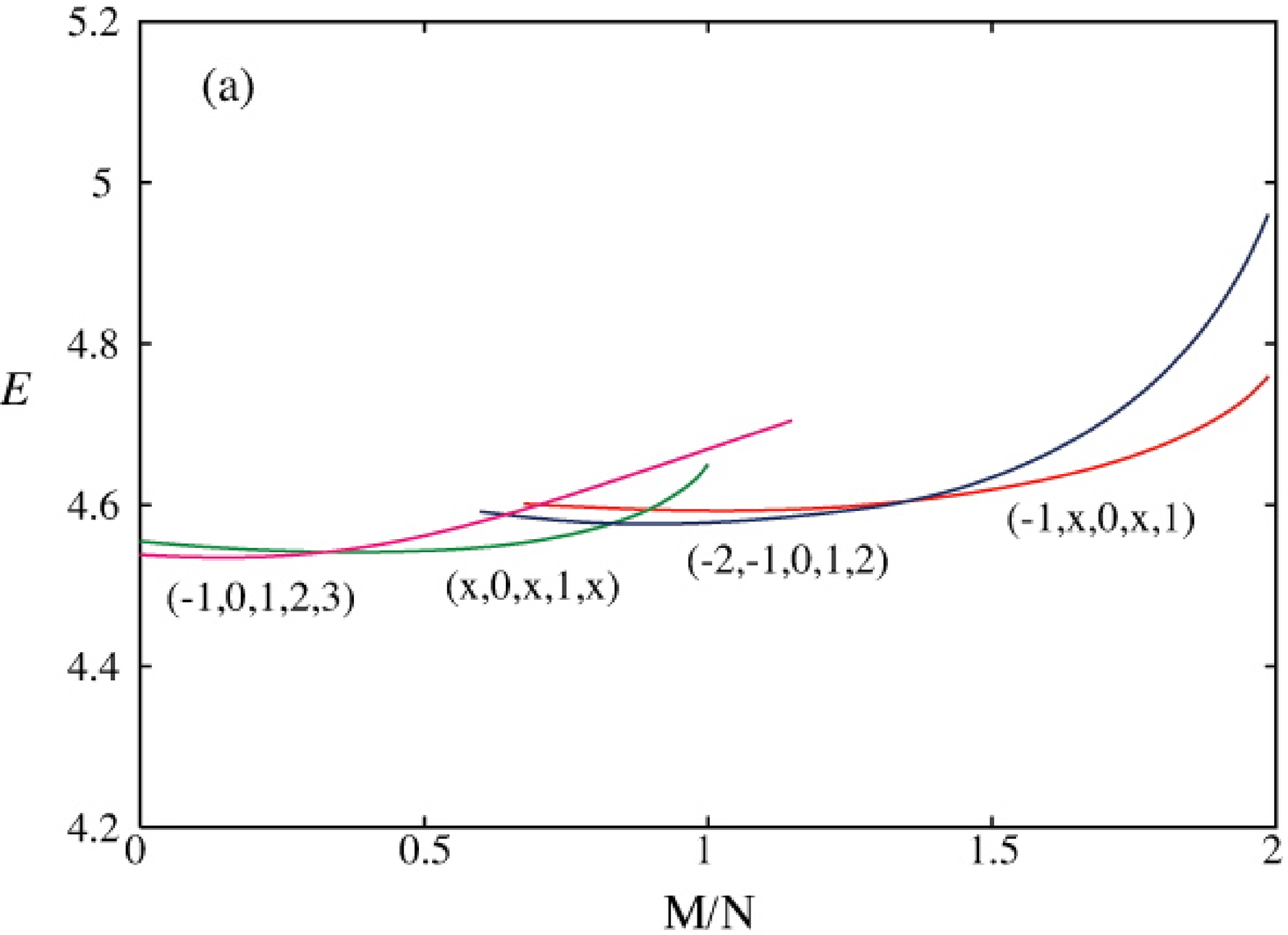} \newline
\includegraphics[width=0.8\linewidth]{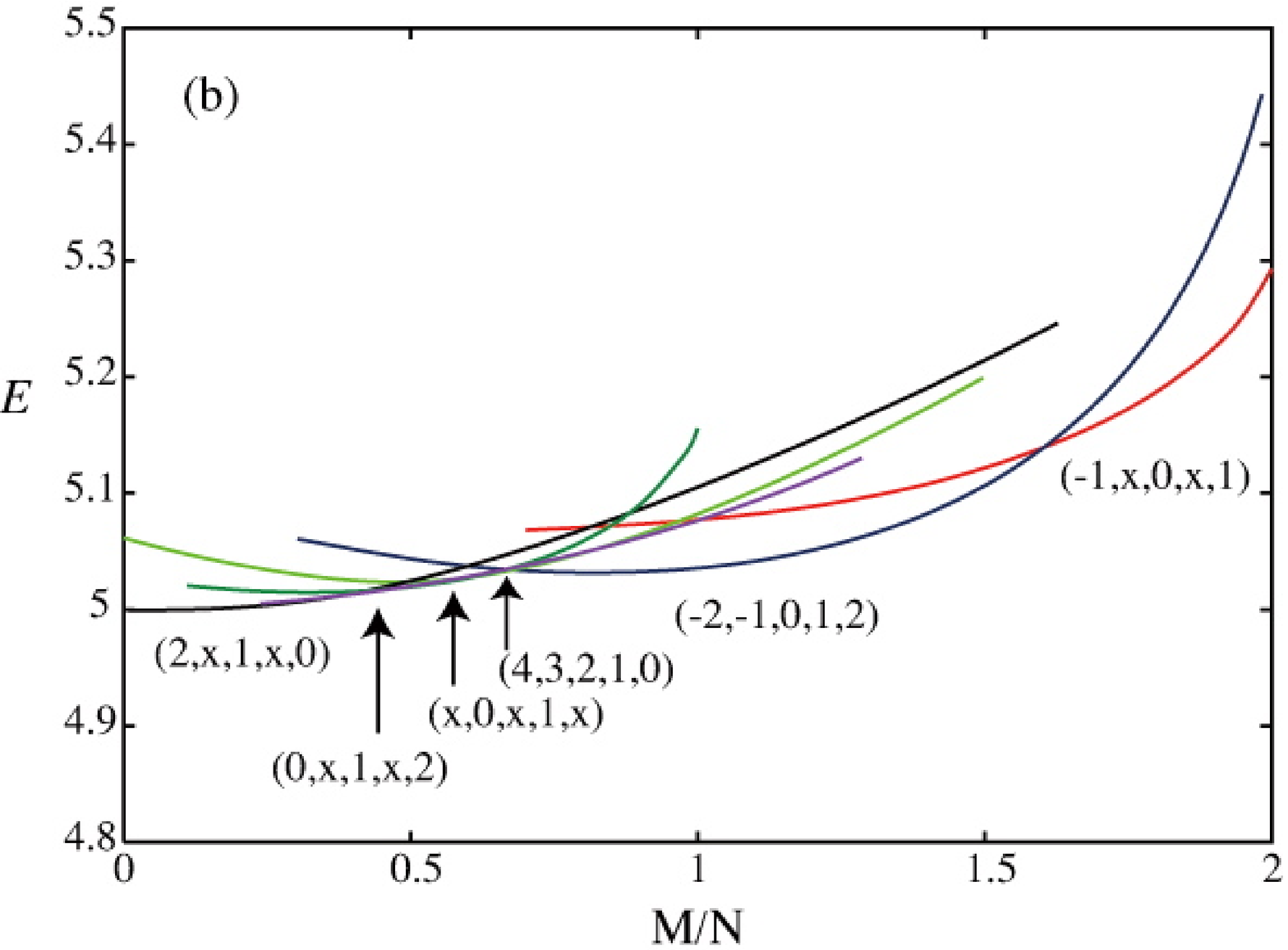} \newline
\includegraphics[width=0.8\linewidth]{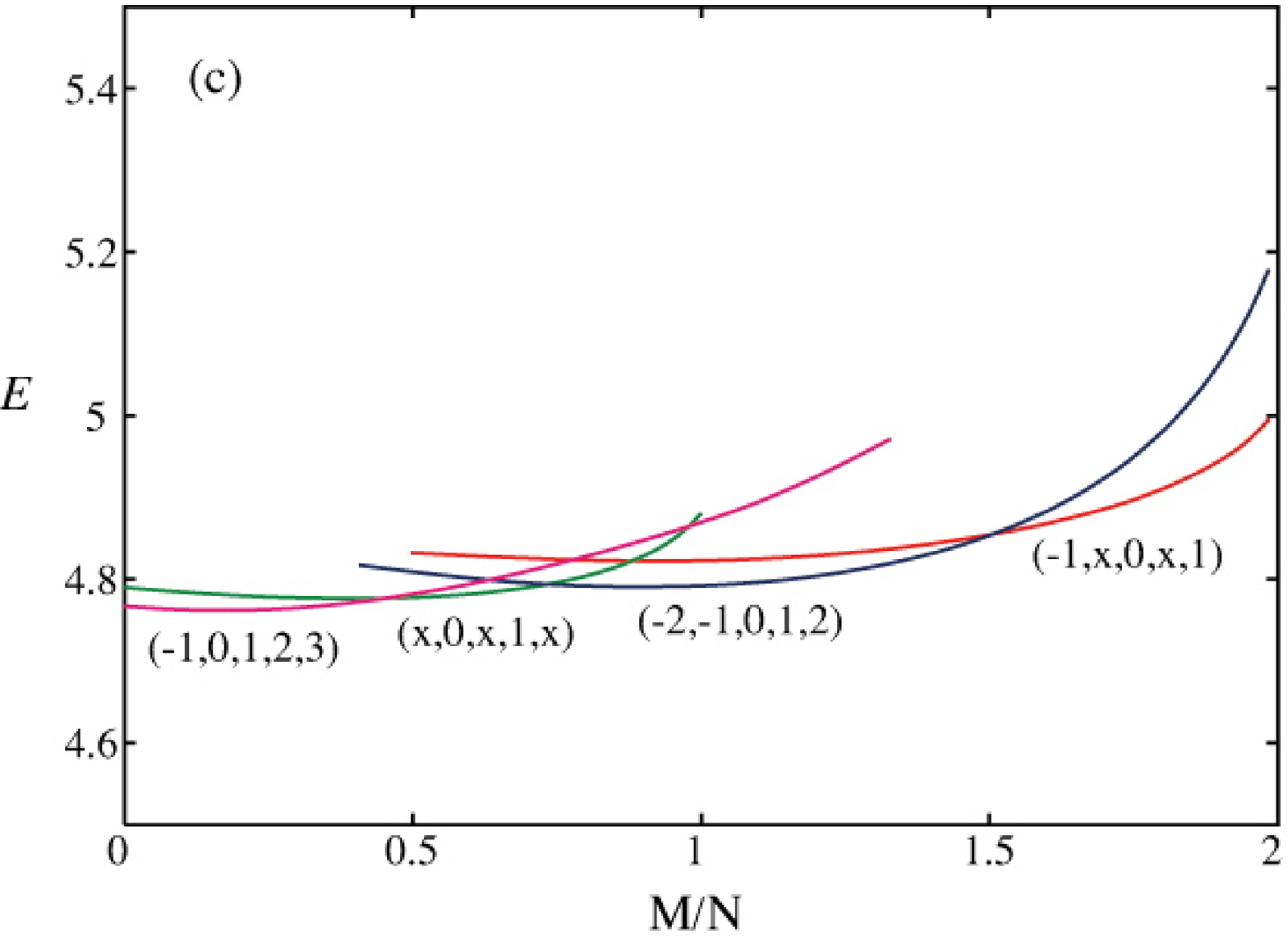}
\caption{ (color online) Dependences of energies of different vortex phases
on the magnetization for cyclic (a), polar (b), and polar+cyclic (c) states.
The energy is measured in units of $\hbar \protect\omega _{\perp }$. }
\label{}
\end{figure}

Which state has the lowest energy at given magnetization, depends on many
factors. For instance, at high magnetization, close to $M/N=2$, condensate
has to be concentrated mostly in the state with $m_{F}=2$. Since our system
is rotated with the frequency enough to create a vortex, for the condensate
it is favorable energetically to have winding number $1$ in this state. For
other particles, which are not in $m_{F}=2$\ state, it is energetically
favorable to be in a superfluid phase with winding number $0$ in order to
occupy the inner part of the trap, where the trapping potential is small.
That is why in all three phase diagrams presented in Fig.~1 vortex phase $%
(-1,\times ,0,\times ,1)$ has the lowest energy at high magnetization. The
dependences of the density of particles in each hyperfine state on the
distance from the potential well center is shown on Fig.~2(a) for the case
of polar phase at $M/N=1.87$. Spatial variations of the order parameters $%
\Theta $\ and $\left\vert \left\langle \mathbf{f}\right\rangle \right\vert $
are presented \ on Fig.~2(b). It can be seen from Fig.~2(b) that $\Theta $
is maximum in the center of the potential well and vanishes at the infinity.
At the same time, $\left\vert \left\langle \mathbf{f}\right\rangle
\right\vert $ has a minimum at $r=0$ and tends to $2$ at $r\rightarrow
\infty $. This is because close to $r=0$\ a component with $m_{F}=0$ has a
maximum and all other components are small. Therefore, $\Theta $, which
characterizes a formation of singlet pairs, has a maximum at $r=0$, and $%
\left\vert \left\langle \mathbf{f}\right\rangle \right\vert =0$. Far from
the center of the cloud, the density of particles with $m_{F}=2$\ is much
larger than that for $m_{F}=0$ and the densities in all other hyperfine
states are very small. For this reason, $\Theta =0$\textbf{\ }and\textbf{\ }$%
\left\vert \left\langle \mathbf{f}\right\rangle \right\vert =2$ at \ $%
r\rightarrow \infty $.

\begin{figure*}[t]
\begin{tabular}{ccc}
\includegraphics[width=0.32\linewidth]{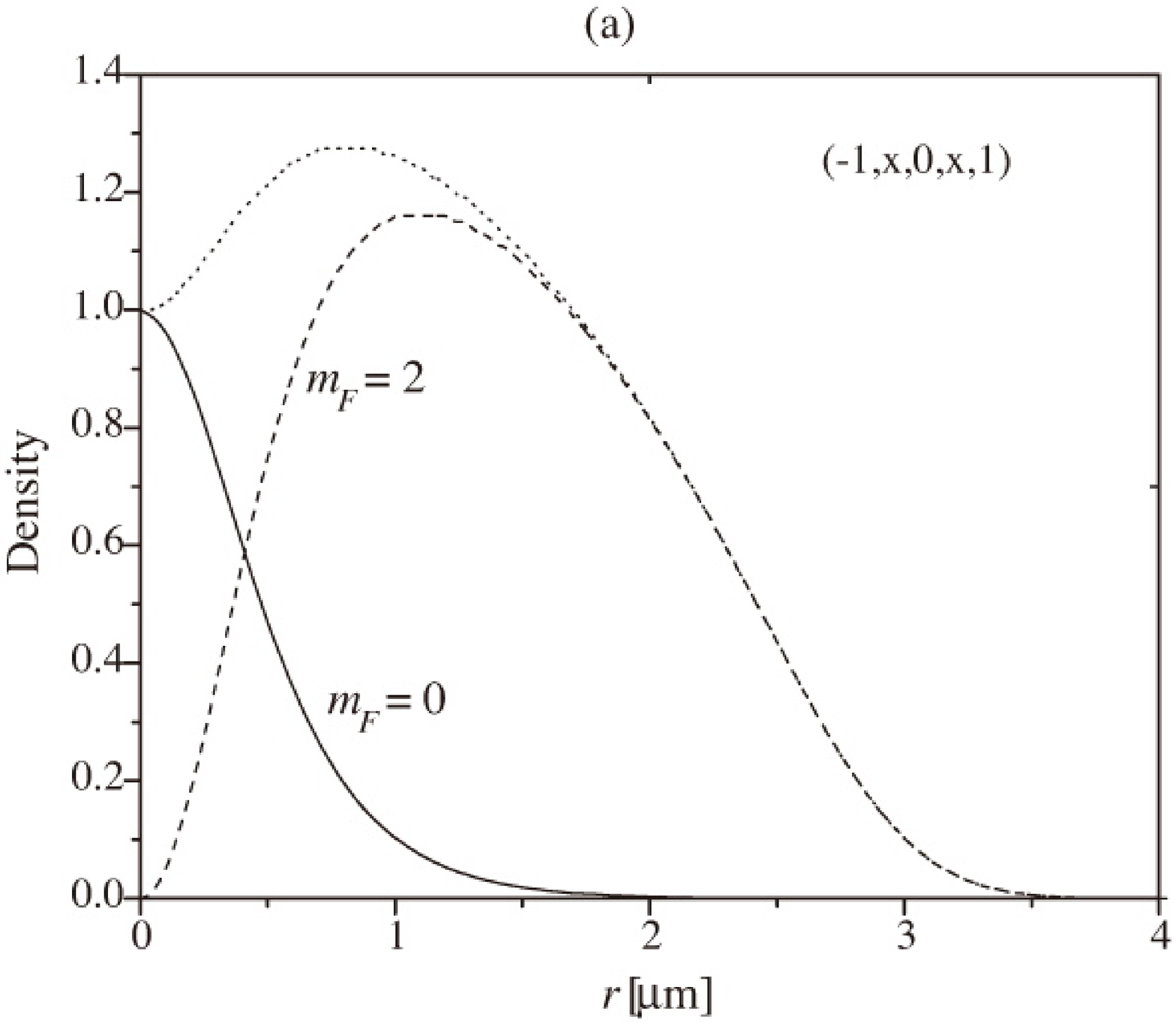} & %
\includegraphics[width=0.32\linewidth]{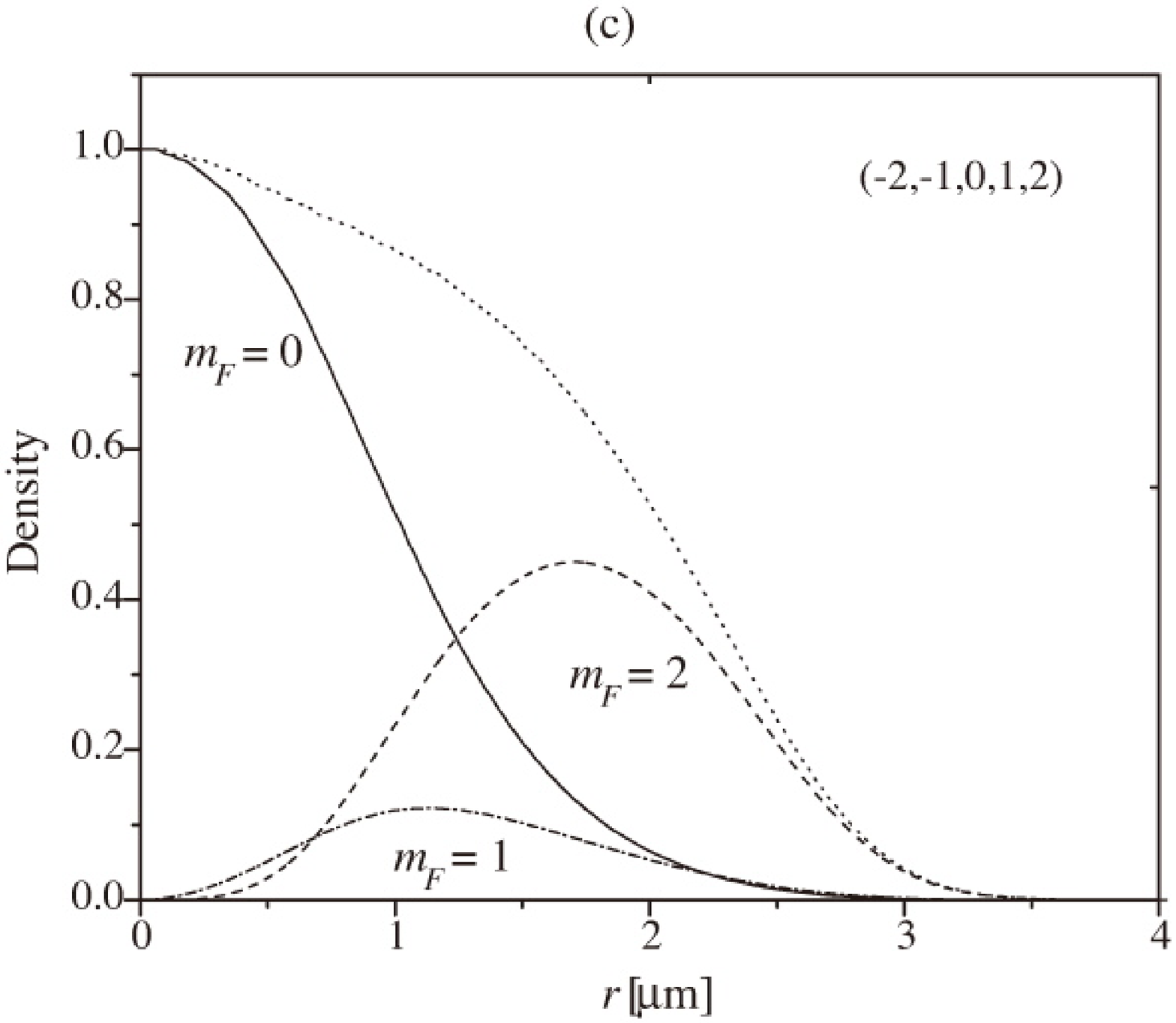} & %
\includegraphics[width=0.32\linewidth]{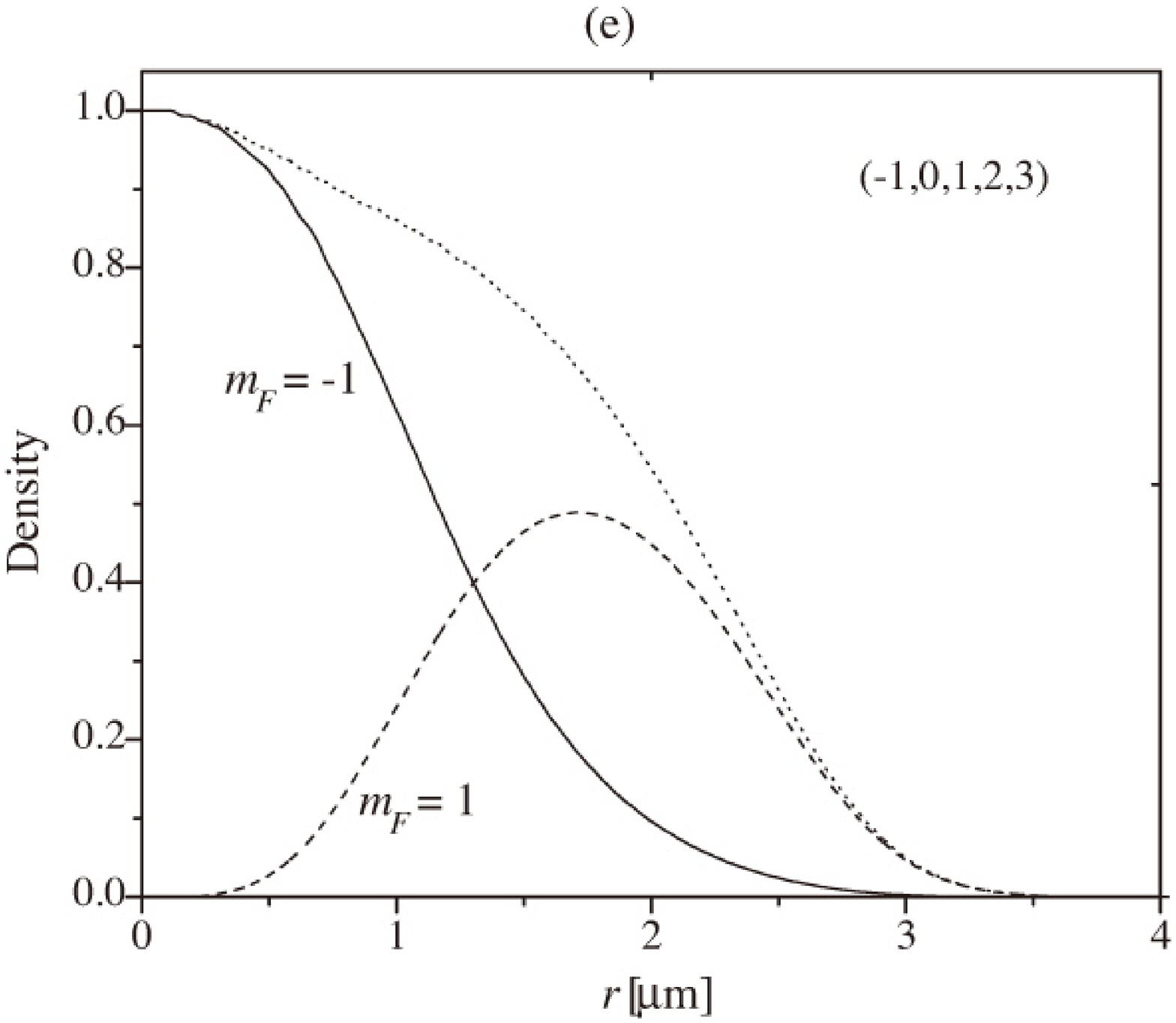} \\ 
\includegraphics[width=0.32\linewidth]{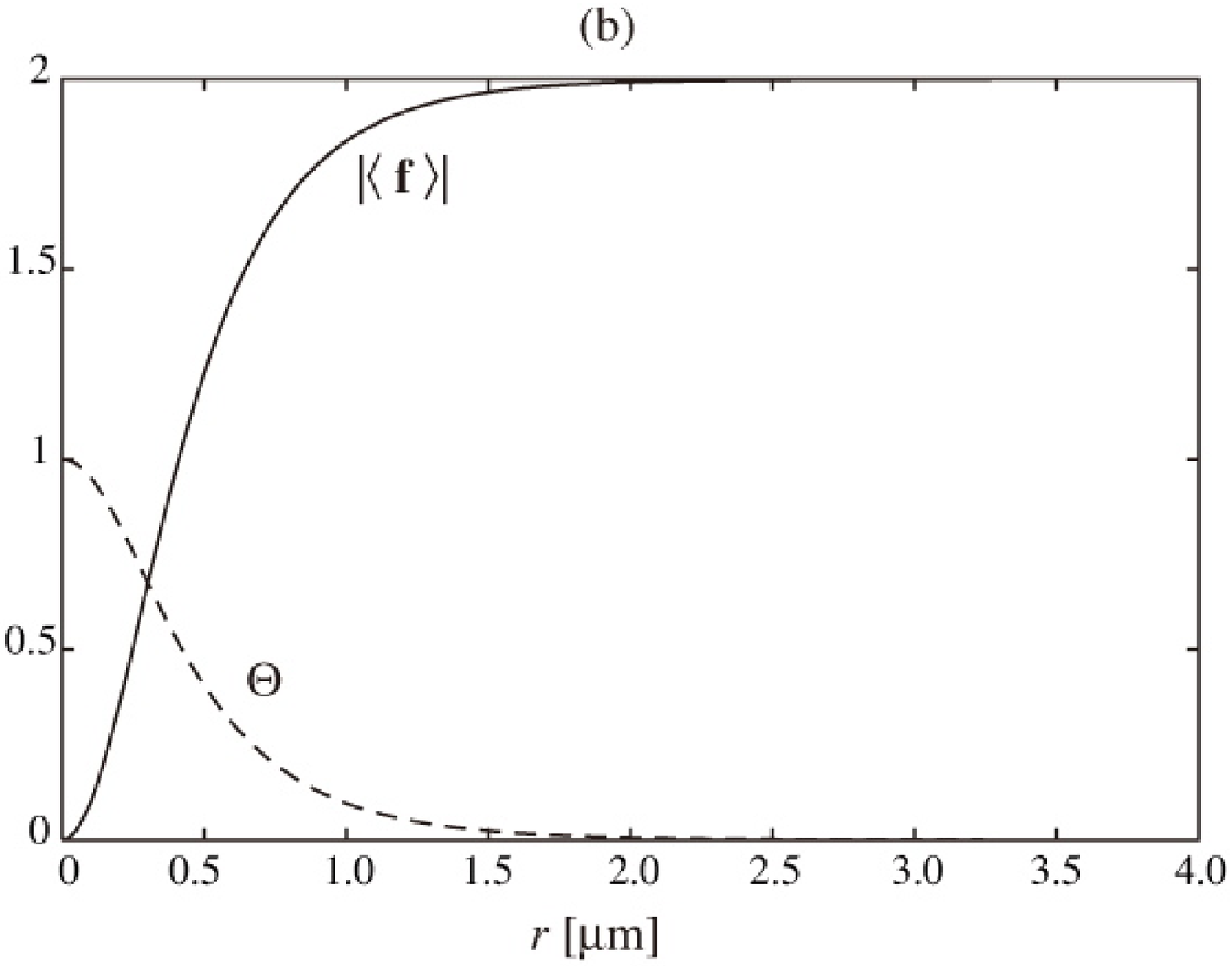} & %
\includegraphics[width=0.32\linewidth]{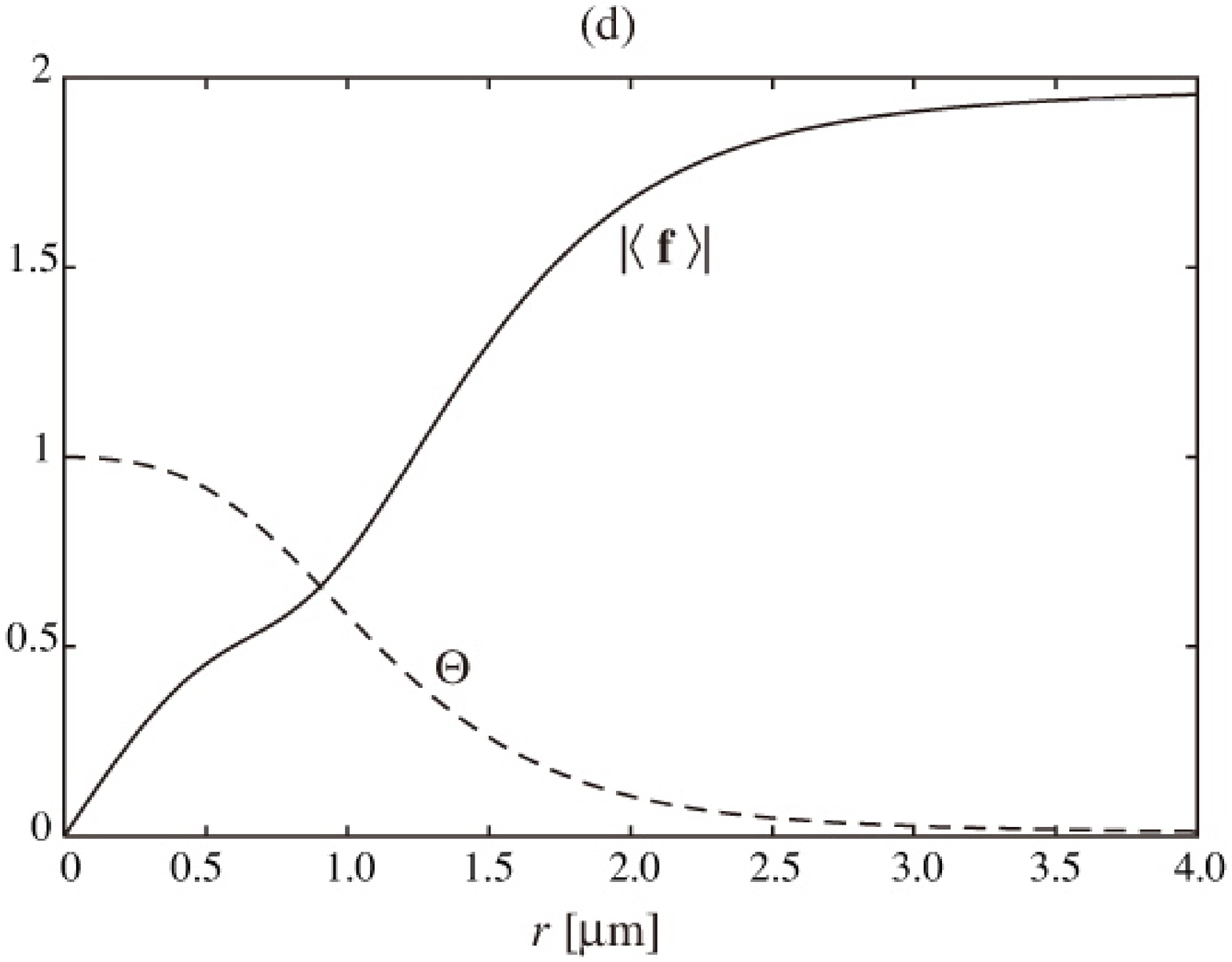} & %
\includegraphics[width=0.32\linewidth]{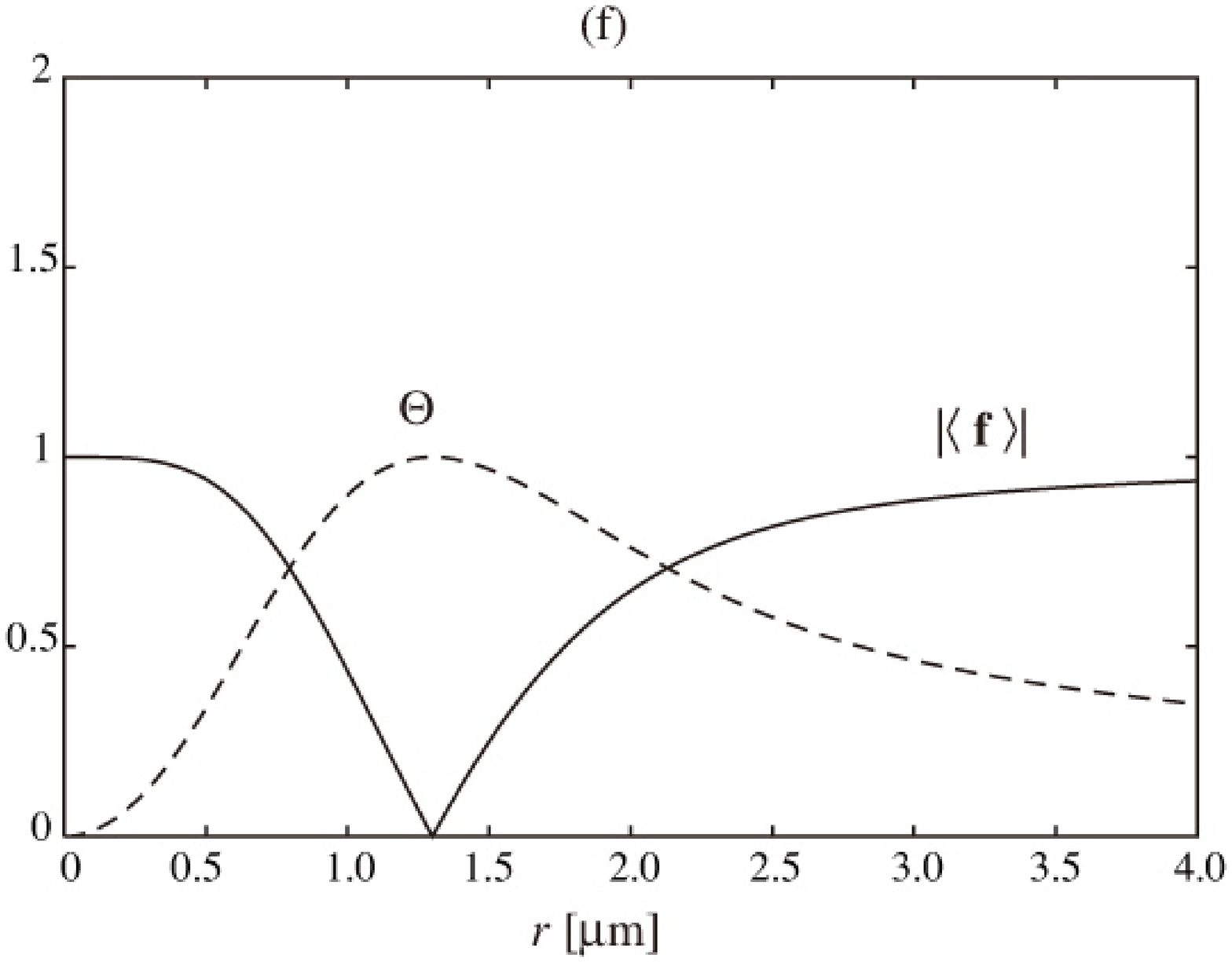}%
\end{tabular}%
\caption{ The spatial variation of the density of particles in different
hyperfine states normalized by the total density at the system axis and the
order parameters $\Theta $\ and $\left\vert \left\langle \mathbf{f}%
\right\rangle \right\vert $ for different vortex phases. Fig. 2 (a) and (b)
correspond to the $(-1,\times ,0,\times ,1)$ polar phase at $M/N=1.87$;
Fig.~2 (c) and (d) to the $(-2,-1,0,1,2)$ phase at $M/N=1.21$; Fig. 2 (e)
and (f) to the cyclic+polar $(-1,0,1,2,3)$ state at $M/N=0.18$. Dot lines
show the total density of particles. Total number of particles is 10000.}
\end{figure*}

For lower magnetization, condensate has to be distributed between the states
with $m_{F}=0,1,2$.\ It turns out that again in all phase diagrams in Fig.~1
the ground state is represented by the phase $(-2,-1,0,1,2)$ in rather broad
range of $M$. This is due to the fact that, in this case, it is favorable to
put most of the particles in the state with winding number $0$ in order to
occupy the inner part of the trap. Most of remaining particles are condensed
to the state with winding number $2$ occupying the outer part of the trap
and thus decreasing the energy of interaction (in "density" channel) of
particles with $m_{F}=0$\ and $2$. Typical profiles of the particles
densities in different hyperfine states for this vortex phase are shown in
Fig.~2(c) for the cyclic state at $M/N=1.21$. Fig.~2(d) indicates $r$%
-dependences of the order parameters $\Theta $\ and $\left\vert \left\langle 
\mathbf{f}\right\rangle \right\vert $, which physically are similar to that
in the case of $(-1,\times ,0,\times ,1)$ state shown in Fig. 2 (a) and (b).

At lower $M$ particles have to be distributed between several states with
different $m_{F}.$\ In this case, spin interactions become important and,
therefore, the sequences of phase transitions for different phase diagrams
in Fig.~1 are different. We can conclude that at high magnetizations, $%
M\gtrsim 1$, the state with lowest energy is mostly determined by
interactions in the "density" channel, whereas at low magnetization $M\sim 0$%
, spin interactions play an important role. In Fig.~2(e) and (f) we present
the $r$-dependences of the superfluid density in different hyperfine states
and order parameters $\Theta $\ and $\left\vert \left\langle \mathbf{f}%
\right\rangle \right\vert $ for the cyclic+polar $(-1,0,1,2,3)$ state at $%
M/N=0.18$, where this phase has the lowest energy. In this case, particles
are distributed between states $m_{F}=-1$ and $m_{F}=1$. The state with $%
m_{F}=-1$ has a winding number $0$, and these particles occupy a space with
minimal trapping potential. All other particles are condensed in the state
with winding number $2$ thus decreasing the interaction energy in the
"density" channel. As a result, the order parameter $\Theta $ is nonzero
everywhere except of the point $r=0$, since at any $r>0$ there are particles
in the states with $m_{F}=\pm 1$, and the formation of singlet pairs is
possible. The order parameter $\left\vert \left\langle \mathbf{f}%
\right\rangle \right\vert $\ is nonzero at $r=0$ and $r\rightarrow \infty $,
since in both cases there are particles condensed in the states with nonzero 
$m_{F}$. At the same time, at small values of $r$ most of particles are
condensed in the state with $m_{F}=-1$ and at larger $r$ in the state with $%
m_{F}=1$. Therefore, there is an abrupt change in the spin direction at
intermediate values of r. This results in the vanishing of $\left\vert
\left\langle \mathbf{f}\right\rangle \right\vert $ near the vortex core.

In Fig.~3 we show the spin texture, $\left\langle f_{x}\right\rangle $ and $%
\left\langle f_{y}\right\rangle $, for the phases $(-1,0,1,2,3)$\ and $%
(-2,-1,0,1,2)$\ in cyclic state. In the first case, in the vortex-core
region, the spins are polarized along the z-direction. In the outer region
the spin vanishes and $\Theta $ grows, where the spin amplitude has a node
and the pure polar state forms. In the second case, in the outer region, the
spins are polarized along the z-direction. In the core-region, the spins
lean toward the origin. At the origin, the spin vanishes and the pure polar
state forms ($\Theta =1$), because the spin texture exhibits the $2$%
-dimensional radial disgyration in the core region. Note that $(-2,-1,0,1,2)$
and $(-1,0,1,2,3)$ vortices were not described before in a literature.

\begin{figure*}[t]
\includegraphics[width=0.35\linewidth]{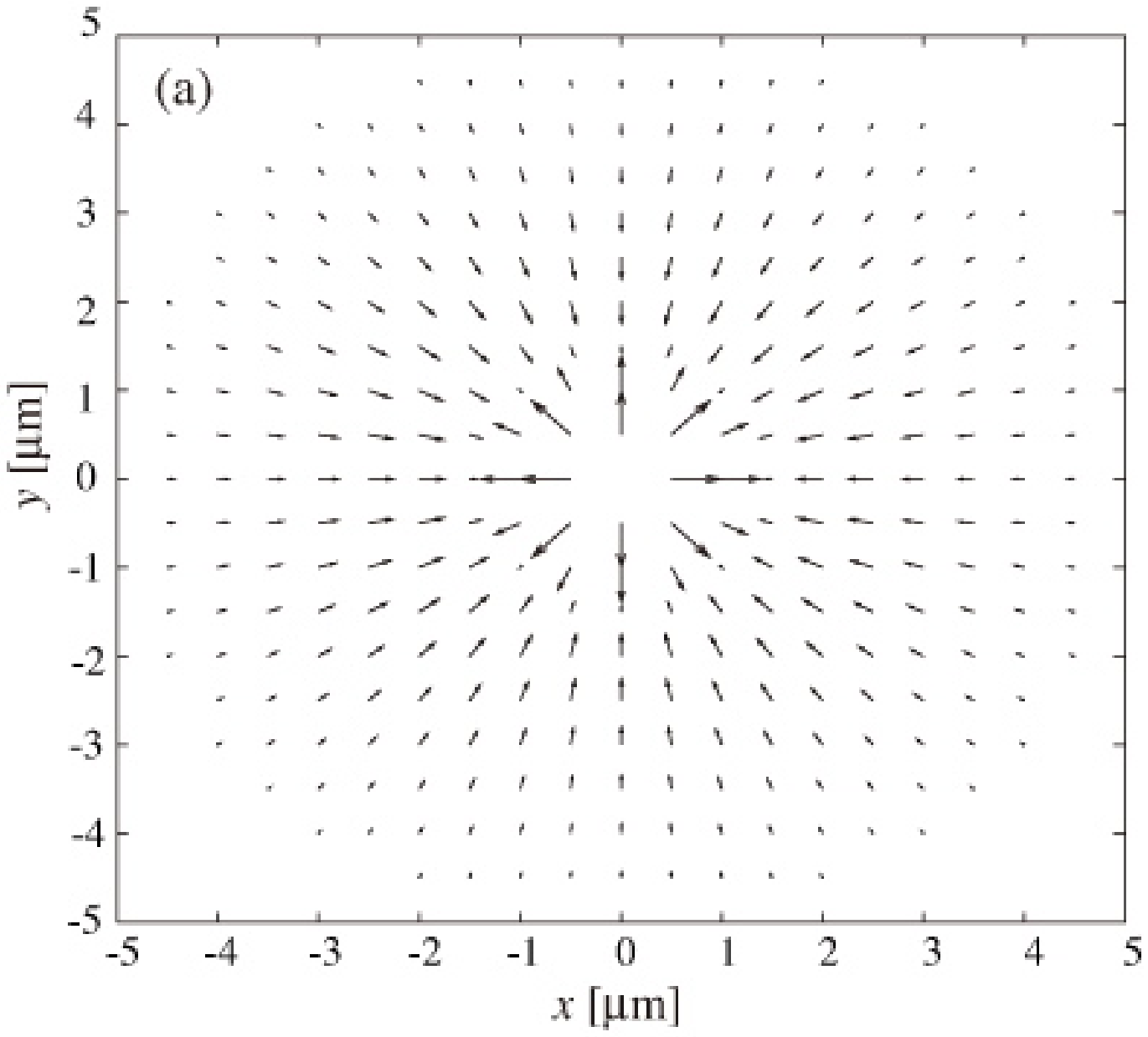} 
\includegraphics[width=0.35\linewidth]{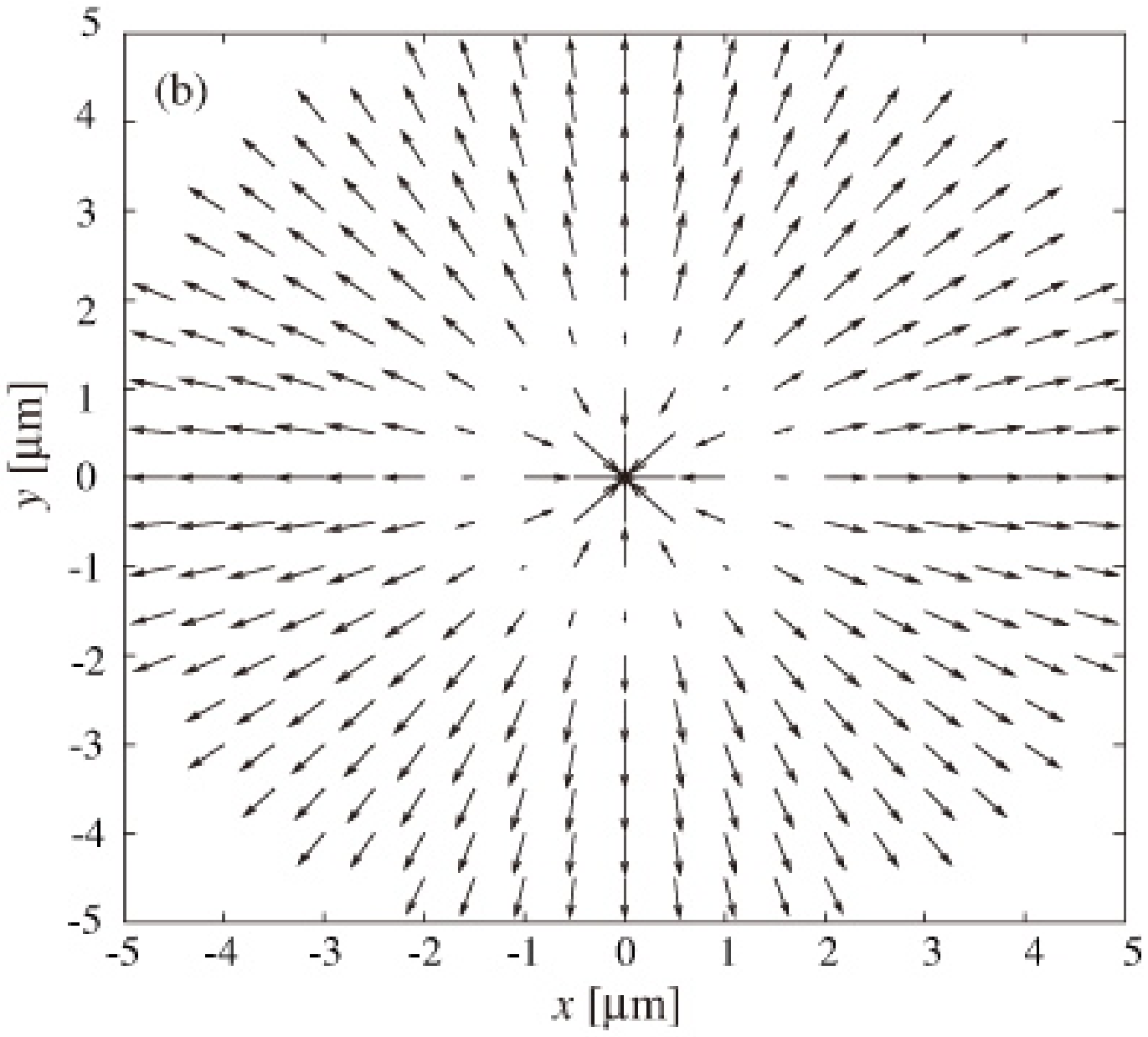}
\caption{ Spin textures, $\left\langle f_{x}\right\rangle $ and $%
\left\langle f_{y}\right\rangle $, for the cyclic $(-1,0,1,2,3)$ phase at $%
M/N=0.16$ (a) and the cyclic $(-2,-1,0,1,2)$ phase at $M/N=1.21$ (b).}
\end{figure*}

\section{Conclusions}

In summary, we analyzed the vortex structure in spinor $F=2$ condensate
using extended Gross-Pitaevskii equations. We considered only
axially-symmetric vortices. Based on symmetric configurations, all possible
vortex states were classified. The Gross-Pitaevskii equations were solved
both numerically and by the variational method using trial functions for the
order parameter. Spatial distribution of the particles density and the order
parameters $\Theta $\ and $\left\vert \left\langle \mathbf{f}\right\rangle
\right\vert $ were obtained. Energies of different vortex phases were found
as a function of magnetization, when the system is rotated with the
frequency $\Omega =0.4\omega _{\perp }$. We found that at high magnetization
the energy of the system is mostly determined by the interaction in the
"density" channel, whereas at low magnetization spin interaction plays an
important role. Also, two new types of vortices were described.

\acknowledgments

W. V. Pogosov is supported by the Japan Society for the Promotion of Science.

%
%
%
%


\end{document}